\documentclass[12pt,cqg]{iopart}
\pdfoutput=1

\usepackage{amsopn}
\usepackage{mathrsfs}
\usepackage{iopams}
\usepackage[numbers,square,sort&compress]{natbib}
\usepackage{graphicx}
\usepackage[pdftex]{hyperref}

\DeclareMathOperator*{\wlim}{w-lim}
\newcommand{\be}{\begin{equation}}
\newcommand{\ee}{\end{equation}}
\newcommand{\ba}{\begin{equation}\begin{aligned}}
\newcommand{\ea}{\end{aligned}\end{equation}}

\DeclareMathAlphabet{\mathpzc}{OT1}{pzc}{m}{it}

\begin{document}

\title{A Simple, Heuristic Derivation of our ``No Backreaction'' Results}

\author{Stephen R. Green$^1$ and Robert M. Wald$^2$}

\address{$^1$ Perimeter Institute for Theoretical Physics \\ 31 Caroline Street North, Waterloo, Ontario N2L 2Y5, Canada}

\address{$^2$ Enrico Fermi Institute and Department of Physics, The University of Chicago\\5640 South Ellis Avenue, Chicago, Illinois 60637, USA}

\eads{\mailto{sgreen@perimeterinstitute.ca} and \mailto{rmwa@uchicago.edu}}

\begin{abstract} 
  We provide a simple discussion of our results on the backreaction
  effects of density inhomogeneities in cosmology, without mentioning
  one-parameter families or weak limits. Emphasis is placed on the
  manner in which ``averaging'' is done and the fact that one is
  solving Einstein's equation. The key assumptions and results that we
  rigorously derived within our original mathematical framework are
  thereby explained in a heuristic way.
\end{abstract}

\maketitle

\section{Introduction}

In a series of papers~\cite{Green:2010qy,Green:2011wc,Green:2013yua,Green:2014aga}, we provided
a rigorous mathematical framework for analyzing the effects of backreaction produced by
small scale inhomogeneities in cosmology. We proved results showing
that no large backreaction effects can be produced by matter inhomogeneities, provided that the
energy density of the matter is positive in all frames. In particular, we proved that at leading order
in our approximation scheme, the effective stress-energy provided by the nonlinear terms in Einstein's
equation must be traceless and have positive energy in all frames, corresponding to 
the backreaction effects of gravitational radiation.

Recently, our work has been criticized by Buchert et
al~\cite{Buchert:2015iva}.  We have responded to these criticisms
in~\cite{Green:2015bma} and see no need to further amplify our
refutation of these criticisms here. Nevertheless, it has become clear
to us that it would be useful to provide a simple, heuristic
discussion of our results, in order to make more clear various aspects
of our work, including (i) the nature of our assumptions, (ii) the
relationship of our procedures to ``averaging,'' (iii) the manner in
which we use Einstein's equation, and (iv) the significance of our
results. In this way, the basic nature of our results can be seen
without invoking technical mathematical procedures, such as the taking
of weak limits. The price paid for this, of course, is a loss of
mathematical precision and rigor; for example, many of our equations
below will involve the use of the relations ``$\sim$'' or
``$\ll$''---which will not be given a precise meaning---and some terms
in various equations will be dropped because they are ``small.''
However, the reader desiring a more precise/rigorous treatment can
simply re-read our original
papers~\cite{Green:2010qy,Green:2011wc,Green:2013yua,Green:2014aga}.
In section III below, we will provide a guide to relating the
heuristic discussion of the present paper to the precise formulation
(using one-parameter families and weak limits) given in our original
papers.

To begin, the situation that we wish to treat is one where the
spacetime metric, which solves the Einstein equation exactly on all
scales, takes the form
\begin{equation}
g_{ab} = g^{(0)}_{ab} + \gamma_{ab}
\end{equation}
where $g^{(0)}_{ab}$ has ``low curvature'' and $\gamma_{ab}$ is
``small,'' but derivatives of $\gamma_{ab}$ may be large, so that the
geodesics and the curvature (and, hence, the associated stress-energy
distribution) of $g_{ab}$ may differ significantly from that of
$g^{(0)}_{ab}$. In particular, it is not assumed that $g_{ab}^{(0)}$
solves the Einstein equation. This should be an excellent description
of the metric of our universe except in the immediate vicinity of
black holes and neutron stars. To make our assumptions about the form
of the metric a bit more precise, it is convenient to introduce a
Riemannian metric $e_{ab}$ and use it to define norms on all
tensors. We assume that
\begin{equation}
\left|g^{(0)}_{ab}\right| \equiv \left[e^{ac} e^{bd} g^{(0)}_{ab} g^{(0)}_{cd}\right]^{1/2} \sim 1
\end{equation}
whereas
\begin{equation}
\left|\gamma_{ab}\right|  \ll 1 \, .
\label{gam1}
\end{equation}
We denote the curvature length scale associated with $g^{(0)}_{ab}$ by $R$, i.e., 
\begin{equation}
\left|{R^{(0)a}}_{bcd}\right| \sim 1/R^2 \, .
\end{equation}
In cosmological applications, $g^{(0)}_{ab}$ would be taken to be a
metric with FLRW symmetry (but not assumed to satisfy the Friedmann
equations) and $R$ would be the Hubble radius,
$R = R_H \sim 5 \, \rm{Gpc}$ (today), but our arguments apply to much
more general situations. For our universe, apart from the immediate
vicinity of strong field objects, $\gamma_{ab}$ would be largest near
the centers of rich clusters of galaxies, where $|\gamma_{ab}|$ can be
as large as $\sim 10^{-4}$.  However, although $\gamma_{ab}$ is
required to be small, $\gamma_{ab}$ is allowed to have large
derivatives. We require that the first derivatives of $\gamma_{ab}$ be
constrained only by\footnote{Equation (\ref{gam2}) will be used to
  justify dropping various terms that arise in integrals over large
  regions, such as occur in going from eq.~(\ref{t01}) to eq.~(\ref{t01a}) below. It is therefore fine if (\ref{gam2}) fails to hold in
  highly localized regions, such as near the surface of a massive
  body.}
\begin{equation}
\left|\gamma_{ab} \nabla_c \gamma_{de}\right|  \ll 1/R \, ,
\label{gam2}
\end{equation}
where $\nabla_a$ denotes the derivative operator associated with $g^{(0)}_{ab}$.
Second derivatives of $\gamma_{ab}$ are entirely unconstrained, so
locally, we may have 
\begin{equation}
\left|\nabla_c \nabla_d \gamma_{ab}\right| \gg 1/R^2
\end{equation}
Thus, the curvature of $g_{ab}$ is allowed to locally be much greater than 
that of $g^{(0)}_{ab}$, as is the case of main interest for cosmology. 
In this situation, ordinary perturbation theory about $g^{(0)}_{ab}$ cannot be directly 
applied to Einstein's equation for $g_{ab}$, since even though $\gamma_{ab}$ itself is small, the
terms involving $\gamma_{ab}$ that appear in Einstein's equation are not small. 

We assume that the matter in the universe is described by a stress-energy tensor $T_{ab}$ that satisfies
the weak energy condition, 
\begin{equation}
T_{ab} t^a t^b \geq 0
\label{wec}
\end{equation}
for all timelike $t^a$. (Here ``timelike'' means with respect to $g_{ab}$,
i.e., $g_{ab} t^a t^b < 0$, although it makes essentially no difference 
whether we use $g_{ab}$ or $g^{(0)}_{ab}$ since 
$|\gamma_{ab}|  \ll 1$.) We assume further that $T_{ab}$ is (essentially) homogeneous on some scale $L$ with
$L \ll R$. By this we mean that $T_{ab}$ can be written as
\begin{equation}
T_{ab} = T^{(0)}_{ab} + \Delta T_{ab}
\end{equation}
where $|T^{(0)}_{ab}| \lesssim 1/R^2$ and $\Delta T_{ab}$ ``averages''
to (nearly\footnote{\label{fn:lin}We refer to small fluctuations in
  $\Delta T_{ab}$ beyond the homogeneity scale as its
  ``long-wavelength part''~\cite{Green:2011wc}. These fluctuations can
  be described by linear perturbation theory, and will be neglected in
  this paper. For further discussion see Sec.~III
  of~\cite{Green:2010qy}, and~\cite{Green:2011wc}.}) zero on scales
large compared with $L$ (even though $\Delta T_{ab}$ may locally be
extremely large compared with $T^{(0)}_{ab}$). In the case of interest
for cosmology, $T^{(0)}_{ab}$ would have FLRW symmetry.

The assumption that $\Delta T_{ab}$ averages to zero on large scales is a key assumption,
so we should further explain both its meaning and our justification for 
making it. First, it is not obvious what one means by
the ``averaging'' of a tensor quantity such as $\Delta T_{ab}$. In a non-flat
spacetime, parallel transport is path dependent, so the values of tensors at different points cannot be meaningfully compared,
as required to give any invariant meaning to an averaging procedure. 
Now, since $g^{(0)}_{ab}$ is locally flat in any region, 
$\mathcal D$, of size $D \ll R$, averaging of tensor fields over such a
region $\mathcal D$ is well defined. However, we do not want to require $D \ll R$.
Furthermore, even if we restricted the size of 
$\mathcal D$ to $D \ll R$, we don't want to simply integrate quantities
over such a region $\mathcal D$ because the introduction of sharp boundaries for 
$\mathcal D$ will produce artifacts 
that we wish to avoid. We therefore will do our ``averaging'' in the following manner:
We choose a region $\mathcal D$ with
$D > L$ and introduce a smooth tensor field 
$f^{ab}$ with support in $\mathcal D$ such that $f^{ab}$ ``varies as slowly as possible'' over $\mathcal D$ compatible with its vanishing outside of $\mathcal D$
and with the curvature of $g^{(0)}_{ab}$. Specifically, for any
region $\mathcal D$ with
$L < D \lesssim R$, we require $f^{ab}$ to be chosen so that\footnote{If we chose $D > R$, we would have to replace the right
side of this equation with $\max |f^{ab}|/R$. However, there is no reason to choose $D > R$.}
\begin{equation}
\max \left|\nabla_c f^{ab}\right|  \lesssim \max \frac{\left|f^{ab}\right|}{D}
\label{slow}
\end{equation}
A more precise statement of our homogeneity requirement on $T_{ab}$ 
is that $\Delta T_{ab}$ be such that for 
any region $\mathcal D$ with $L < D \lesssim R$ and any such $f^{ab}$, we have
\begin{equation}
\left|\int f^{ab} \Delta T_{ab} \right| \ll \left|\int f^{ab} T^{(0)}_{ab}\right| \lesssim \frac{1}{R^2} \int \left|f^{ab}\right| \, .
\label{avg}
\end{equation}

The integral appearing in eq.~(\ref{avg}) is a spacetime integral over
the region $\mathcal D$. In a general context---where significant
amounts of gravitational radiation may be present and the motion of
matter may be highly relativistic---both $\gamma_{ab}$ and $T_{ab}$
may vary rapidly in both space and time, and it is important that
$\mathcal D$ be sufficiently large in both space and time. However,
for cosmological applications, the case of greatest interest is one in
which there is rapid spatial variation on scales small compared with
the Hubble radius, but time variations are negligibly small. In this
case, it is important that the spatial extent, $D$, of $\mathcal D$ be
larger than the spatial homogeneity scale $L$, but the time extent of
$\mathcal D$ may be taken to be significantly smaller than $D/c$. For
our universe, the assumptions of the previous paragraph should hold
for $L \sim 100 \, {\rm Mpc}$. [Beyond this scale, eq.~(\ref{avg}) does not preclude the presence of small fluctuations in $\Delta T_{ab}$ (see footnote~\ref{fn:lin}).]

To summarize, there are $3$ length scales that appear in our analysis. The first is the
curvature length scale, $R$, of $g^{(0)}_{ab}$ (i.e., the Hubble radius). The second is the homogeneity length scale,
$L$. It is an essential assumption that $L \ll R$. The third is the averaging length scale, $D$, which is up to 
us to choose. We must always choose $D > L$ if we wish to (very nearly) average out 
the stress-energy inhomogeneities. It is never useful to choose $D > R$, since we don't wish to average
over the background structure. For some computations, it will be useful
to choose $D \sim R$, and for others it will be more useful to choose $L < D \ll R$. In all cases, the averaging
will be done over a region $\mathcal D$ of size $D$
using a slowly varying test tensor field [see eq.~(\ref{slow})].

We may interpret $g^{(0)}_{ab}$ as the ``averaged metric'' (although no actual averaging
need be done since $|\gamma_{ab}| \ll 1$), whereas $T^{(0)}_{ab}$ represents the large-scale
average of $T_{ab}$. The issue at hand in whether the small scale inhomogeneities of
$g_{ab}$ and $T_{ab}$ can contribute nontrivially to the dynamics of
$g^{(0)}_{ab}$. A priori, this is possible because even though
$\gamma_{ab}$ is assumed to be small,
Einstein's equation for $g_{ab}$ contains derivatives of
$\gamma_{ab}$, which need not be small. 
Consequently, the average of the Einstein tensor, $G_{ab}$, of $g_{ab}$ need
not be close to the Einstein tensor, $G^{(0)}_{ab}$, of
$g^{(0)}_{ab}$. Thus, although $g_{ab}$ is a assumed to be an exact
solution of Einstein's equation (with cosmological constant, $\Lambda$) with stress-energy source
$T_{ab}$, it is possible that $g^{(0)}_{ab}$ may not be close to a
solution to Einstein's equation with
source $T^{(0)}_{ab}$. If we have
\begin{equation}
G^{(0)}_{ab} + \Lambda g_{ab}^{(0)} - 8 \pi T^{(0)}_{ab} \ll 1/R^2
\end{equation}
then we say that there is a negligible {\em backreaction} effect of the small scale
inhomogeneities on the effective dynamics of $g^{(0)}_{ab}$. Conversely, if
\begin{equation}
G^{(0)}_{ab} + \Lambda g_{ab}^{(0)} - 8 \pi T^{(0)}_{ab} \sim 1/R^2
\end{equation}
then the backreaction effects are large. Our aim is to determine whether the backreaction
effects can be large and, if so, to determine the properties of the averaged 
effective stress-energy tensor of backreaction, defined by
\begin{equation}
8 \pi t^{(0)}_{ab} \equiv G^{(0)}_{ab} + \Lambda g_{ab}^{(0)} - 8 \pi T^{(0)}_{ab} \, .
\end{equation}
The most interesting possibility would be to have large backreaction effects with
$t^{(0)}_{ab}$ of the form $ - C g^{(0)}_{ab}$ with $C \sim 1/R^2$, in which case
the backreaction effects
of small scale inhomogeneities would mimic that of a cosmological constant, 
and the observed acceleration
of our universe could be attributed to these backreaction effects, without the need to postulate
the presence of a true cosmological constant, $\Lambda$, in Einstein's equation. However, we will show that
this is not possible.

Our strategy, now, is simply the following. We write down the exact Einstein equation,
\begin{equation}
G_{ab} + \Lambda g_{ab} = 8 \pi T_{ab} \, .
\label{ee1}
\end{equation}
We then take suitable averages of this equation in the manner described above to obtain an
expression for the effective stress-energy of backreaction, $t^{(0)}_{ab}$, and determine its properties,
using only our assumptions
(\ref{gam1}), (\ref{gam2}), (\ref{wec}), and (\ref{avg}).
In order to implement our strategy, it is extremely useful to write the exact Einstein equation (\ref{ee1}) in the form
\begin{equation}
G^{(0)}_{ab} + \Lambda g_{ab}^{(0)} - 8 \pi T^{(0)}_{ab} = 8 \pi \Delta T_{ab} - \Lambda \gamma_{ab} - G^{(1)}_{ab} - G^{(2)}_{ab} - G^{(3+)}_{ab} \, .
\label{ee2}
\end{equation}
Here $G^{(1)}_{ab}$ denotes the terms in the exact Einstein tensor $G_{ab}$
that are linear in $\gamma_{ab}$; $G^{(2)}_{ab}$ denotes the terms in $G_{ab}$
that are quadratic in $\gamma_{ab}$; and $G^{(3+)}_{ab}$ denotes the terms in $G_{ab}$
that are cubic and higher order in $\gamma_{ab}$. 

Before proceeding with our analysis, we comment that if, 
following~\cite{Ishibashi:2005sj}, one simply inserts Newtonian estimates for $\gamma_{ab}$
associated with the various 
density inhomogeneities found in our universe (clusters of galaxies, galaxies, stars,
etc.), one can easily see that the backreaction effects in our universe are negligible.
Our aim here is to significantly improve upon such ``back of the envelope estimates''
by showing that the backreaction associated with density inhomogeneities can
\emph{never} be large if (\ref{gam1}), (\ref{gam2}), (\ref{wec}), (\ref{avg}), and \eref{ee1} hold. 

\section{Determination of the Averaged Effective Stress-Energy}

We now analyze the contributions of the various terms on the right side of eq.~(\ref{ee2}).
Our first claim is that, under our assumptions, the contribution of $G^{(3+)}_{ab}$ is negligible
compared with $G^{(2)}_{ab}$.
This is because no term in Einstein's equation contains more than a total of $2$ derivatives. Thus,
a term that is cubic or higher order in $\gamma_{ab}$ must contain at least one factor of
$\gamma_{ab}$ that is undifferentiated. Since $|\gamma_{ab}| \ll 1$, all such terms will be
much smaller than corresponding terms in $G^{(2)}_{ab}$. Thus, we neglect $G^{(3+)}_{ab}$
in eq.~(\ref{ee2}). As far as we are aware,
this conclusion is in agreement with all other approaches to backreaction,
i.e., we are not aware of any approach to backreaction that claims that the dominant effects
are produced by cubic or higher order terms in Einstein's equation.

We now average Einstein's equation (\ref{ee2}) (with $G^{(3+)}_{ab}$
discarded) in the manner described in the previous section: We choose
a region $\mathcal D$ of size\footnote{If time variations are slow, we
  may choose the time extent of $\mathcal D$ to be smaller than
  $R/c$.}  $D \sim R$, choose a slowly varying $f^{ab}$ with support
in $\mathcal D$ [see eq.~(\ref{slow})], multiply eq.~(\ref{ee2}) by $f^{ab}$, and integrate.
The term $\int f^{ab} \Delta T_{ab}$ may neglected by
eq.~(\ref{avg}). We estimate $\int f^{ab} G^{(1)}_{ab}$ by integrating
by parts to remove all derivatives\footnote{If the time derivatives of
  $\gamma_{ab}$ are negligibly small, there is no need to integrate by
  parts on the time derivatives, which is why the time extent of
  $\mathcal D$ may be chosen to be smaller than its spatial extent.}
from $\gamma_{ab}$. We obtain
\begin{equation}
\fl \left|\int f^{ab} G^{(1)}_{ab} \right| \lesssim \int |\nabla_c \nabla_d f^{ab}| |\gamma_{ab}|
\lesssim \frac{1}{R^2} \max |\gamma_{ab}| \int |f^{ab}| \ll  \frac{1}{R^2} \int |f^{ab}| .
\end{equation}
Therefore, the contribution of $G^{(1)}_{ab}$ to the averaged Einstein
equation may be neglected. Similarly, the contribution from
$\Lambda \int f^{ab} \gamma_{ab}$ may be neglected, and to our level of approximation we obtain
\begin{equation}
\int f^{ab} t^{(0)}_{ab} = - \frac{1}{8 \pi} \int f^{ab} G^{(2)}_{ab} \, .
\label{t01}
\end{equation}
Furthermore, the terms in $G^{(2)}_{ab}$ can be divided into two types: (i) terms 
quadratic in first derivatives of $\gamma_{ab}$, i.e., of the form
$(\nabla \gamma)(\nabla \gamma)$, and (ii) terms of the form $\gamma \nabla \nabla \gamma$.
For the terms in category (ii), we integrate by parts on one of the derivatives of $\gamma$
to eliminate the second derivative terms. This 
derivative then will either act on the other $\gamma$ factor---thereby converting it to a term of type
(i)---or it will act on $f^{ab}$---in which case it can be neglected on account of eqs.~(\ref{gam2}) and (\ref{slow}). Consequently, we may replace $G^{(2)}_{ab}$ in eq.~(\ref{t01})
by an expression, ${\widetilde G}^{(2)}_{ab}$,
that is quadratic in first
derivatives of $\gamma_{ab}$, and we may rewrite 
eq.~(\ref{t01}) as
\begin{equation}
\int f^{ab} t^{(0)}_{ab} = - \frac{1}{8 \pi} \int f^{ab} {\widetilde G}^{(2)}_{ab} \, .
\label{t01a}
\end{equation}
for all $f^{ab}$ that are ``slowing varying'' in the
sense discussed above.
In order to make contact with commonly used terminology and notation (at least in discussions of gravitational
radiation), it is useful to note that eq.~(\ref{t01a}) can be rewritten as
\begin{equation}
t^{(0)}_{ab} = - \frac{1}{8 \pi} \langle G^{(2)}_{ab} \rangle \, .
\label{t02}
\end{equation}
where $\langle G^{(2)}_{ab} \rangle$ denotes the ``Isaacson 
average''~\cite{Isaacson:1967zz,Isaacson:1968zza} of $G^{(2)}_{ab}$, i.e., 
the quantity obtained by replacing 
$G^{(2)}_{ab}$ by ${\widetilde G}^{(2)}_{ab}$ and ``averaging'' over a region 
of size $D > L$. 
The meaning of eq.~(\ref{t02}) is, of course, simply
that eq.~(\ref{t01a}) holds for all $f^{ab}$ that are ``slowing varying'' in the
sense discussed above.

At this stage, we have ``averaged'' the Einstein equation subject to
our assumptions on this sizes of various terms. The average of the
Einstein equation, however, contains much less information than the
full Einstein equation, which holds at each spacetime point. This
ignorance is encapsulated in an effective backreaction stress-energy
tensor $t_{ab}^{(0)}$ that is completely unconstrained, except to be
expressed as an average of terms quadratic in first derivatives of
$\gamma_{ab}$. Further constraints on $t_{ab}^{(0)}$ can, however, be
obtained by using the fact that the Einstein equation must hold at
each spacetime point.

Our main results~\cite{Green:2010qy} on $t^{(0)}_{ab}$ are that it is
traceless and that it satisfies the weak energy condition. The proof
of these results requires some complicated calculations that were done
in~\cite{Green:2010qy}. Rather than repeat these calculations here, we
will simply outline the logic of our arguments within the heuristic
framework of the present paper, referring the reader
to~\cite{Green:2010qy} for the details of the various calculations.

The quantity ${\widetilde G}^{(2)}_{ab}$ is given by a rather
complicated expression, and in order to make progress on determining
its properties, we need additional information about
$\gamma_{ab}$. However, the only available information about
$\gamma_{ab}$ comes from Einstein's equation (\ref{ee2}). On
examination of this equation, one sees that, locally, the potentially
largest terms are $8 \pi \Delta T_{ab}$ and $G^{(1)}_{ab}$. Therefore,
it might be tempting to set these potentially largest terms to zero by
themselves, i.e., to postulate that the equation
$G^{(1)}_{ab} = 8 \pi \Delta T_{ab}$ holds. Indeed, equations along
these lines (with certain gauge choices) were imposed
in~\cite{Isaacson:1967zz,Isaacson:1968zza,MTW}.  However, as we
explained in section III of~\cite{Green:2010qy}, this equation is not
justified. Indeed, if
$G^{(0)}_{ab} + \Lambda g^{(0)}_{ab} \neq 8 \pi T^{(0)}_{ab}$ this
equation is not even gauge invariant.

Nevertheless, we can obtain very useful information about $\langle G^{(2)}_{ab} \rangle$
from Einstein's equation in a completely reliable way
by the following procedure due to Burnett~\cite{Burnett:1989gp}: We multiply
eq.~(\ref{ee2}) by $\gamma_{cd}$ and ``average'' the resulting equation, i.e.,
we multiply the resulting $4$-index tensor equation by a slowing
varying tensor field $f^{cdab}$ with support in a region $\mathcal D$ (with $D \sim R$
as above) and integrate. Since $|\gamma_{ab}| \ll 1$, the only terms in the resulting
equation that are not a priori negligible are the ones arising from 
$8 \pi \Delta T_{ab}$ and $G^{(1)}_{ab}$. We therefore obtain
\begin{equation}
\int f^{cdab} \gamma_{cd} G^{(1)}_{ab} = 8 \pi \int f^{cdab} \gamma_{cd} \Delta T_{ab}  \, .
\label{linee}
\end{equation}

We now show that the right side of eq.~(\ref{linee}) is negligibly small as a consequence of the
weak energy condition on $T_{ab}$. To see this, we choose
$f^{cdab}$ to be of the form $f^{cdab} = f^{cd} t^a t^b$, where $f^{cd}$ and $t^a$ are slowly
varying and $t^a$ is unit timelike. For such an $f^{cdab}$, the right side of eq.~(\ref{linee}) 
becomes 
\begin{equation}
 \int f^{cd} \gamma_{cd} t^a t^b \Delta T_{ab} = \int f^{cd} \gamma_{cd} [\rho - \rho^{(0)}] \, .
\label{linee2}
\end{equation}
where $\rho \equiv T_{ab} t^a t^b$ and $\rho^{(0)} \equiv T^{(0)}
_{ab} t^a t^b$.  Since $|T^{(0)} _{ab}| \lesssim 1/R^2$ and
$|\gamma_{ab}| \ll 1$, the contribution of the $\rho^{(0)}$ term is
negligible. However, since locally we can have $\rho \gg 1/R^2$, it is
conceivable that the $\rho$ term could make a large
contribution. The key point is that positivity of $\rho$ precludes
this possibility because
\begin{eqnarray}
\left|\int f^{cd} \gamma_{cd} \rho\right|  &\leq& \max |\gamma_{cd}| \int |f^{cd}| \rho
\sim  \max |\gamma_{cd}| \int |f^{cd}| \rho^{(0)} \nonumber \\ 
&\sim& \frac{\max |\gamma_{cd}|}{R^2} \int |f^{cd}| \ll \frac{1}{R^2} \int |f^{cd}| \, .
\label{wec2}
\end{eqnarray}
Here the positivity of $\rho$ was used to omit an absolute value sign on $\rho$ in the first
inequality; we were then able to replace $\rho$ by $\rho^{(0)}$ in the next (approximate)
equality because $|f^{cd}|$ is slowly varying. By contrast, if $\rho$ were allowed to have large
fluctuations of both positive and negative type, then we would not be able to get an
estimate similar to (\ref{wec2}). Basically, for $\rho \geq 0$, although we may have arbitrarily
large positive density fluctuations in localized regions, we must compensate for these by having 
large voids where the density fluctuations are only mildly negative. The net contribution
to eq.~(\ref{linee2}) is then negligible.

Thus, we have shown that the right side of eq.~(\ref{linee}) may be neglected when 
$f^{cdab}$ is of the form $f^{cd} t^a t^b$ for any (slowly varying) $f^{ab}$ and any
(slowing varying) unit timelike $t^a$. But any slowly varying $f^{cdab}$ can be approximated
by a linear combination of terms of the form $f^{cd} t^a t^b$ (with different choices of 
$t^a$ as well as $f^{ab}$). It follows that the right side of eq.~(\ref{linee}) is negligible for all
slowly varying $f^{cdab}$, as we desired to show, and hence
\begin{equation}
\int f^{cdab} \gamma_{cd} G^{(1)}_{ab} = 0 \, .
\label{linee3}
\end{equation}

The expression for $G^{(1)}$ is of the form $\nabla \nabla \gamma$. We can again
integrate by parts in eq.~(\ref{linee3}) to remove one of these derivatives from $\gamma$.
This derivative then will either act on the other $\gamma$ factor or it will act on $f^{abcd}$, in which case it can be neglected. Thus, we may rewrite eq.~(\ref{linee3}) as
\begin{equation}
\langle \gamma_{cd} G^{(1)}_{ab} \rangle = 0 
\label{t04}
\end{equation}
where the ``Isaacson average'' again denotes the average of the quantity quadratic 
in first derivatives of $\gamma$ obtained by integration by parts in eq.~(\ref{linee3}).

In view of eq.~(\ref{t02}), the tracelessness of $t^{(0)}_{ab}$ is
then an immediate consequence of eq.~(\ref{t04}) together with the mathematical fact that
\begin{equation}
\langle  {G^{(2)a}}_a \rangle = \frac{1}{2} \langle \gamma^{ab} G^{(1)}_{ab} \rangle \, ,
\end{equation}
as can be seen by direct inspection of the explicit formulas for both sides of this equation.
We refer the reader to~\cite{Green:2010qy} or~\cite{Burnett:1989gp} for the details. 

The demonstration that $t^{(0)}_{ab}$ satisfies the weak energy condition---i.e., that
$t^{(0)}_{ab} t^a t^b \geq 0$---is considerably more difficult. To show this, it is convenient
to now work in an ``averaging region'' $\mathcal D$ with $L < D \ll R$. We 
choose a point $P \in \mathcal D$ and a unit timelike vector $t^a$ at $P$ 
and construct Riemannian
normal coordinates (with respect to $g^{(0)}_{ab}$) starting from $P$. Since $D \ll R$,
these coordinates will cover $\mathcal D$ and the components of $g^{(0)}_{ab}$ will 
take a nearly Minkowskian form in $\mathcal D$. We choose a positive function $f$
with support in $\mathcal D$ that is slowly varying in the sense that we have 
been using above [see eq.~(\ref{slow})]. Our aim is to show that for any such $f$ we have
\begin{equation}
\int f t^{(0)}_{ab} t^a t^b \geq 0
\label{f}
\end{equation}
where $t^a$ has been extended to $\mathcal D$ via our Riemannian normal coordinates,
i.e., $t^a = (\partial/\partial t)^a$. 

Following~\cite{Green:2010qy}, we start with formula (\ref{t02}) for $t^{(0)}_{ab}$. By 
some relatively nontrivial manipulations using eq.~(\ref{t04}), it turns out that it is possible
to rewrite $\langle G^{(2)}_{ab} \rangle t^a t^b$ entirely in terms of spatial derivatives of spatial components
of $\gamma_{ab}$. The desired formula, derived in~\cite{Green:2010qy}, is
\begin{equation}
\fl \int f t^{(0)}_{ab} t^a t^b = \frac{1}{32 \pi} \int d^4 x f \left[ \partial_i \gamma_{jk} \partial^i \gamma^{jk}
- 2 \partial_j \gamma_{ik} \partial^i \gamma^{jk} +2 \partial_j {\gamma_i}^i \partial_k \gamma^{jk}
-\partial_i {\gamma_j}^j \partial^i {\gamma_k}^k \right]
\label{t05}
\end{equation}
where $i,j,k$ run over the spatial indices of the Riemannian normal coordinates, $\partial_i$
denotes the partial derivative operator in these coordinates, and the raising and lowering of
indices is done using the (essentially flat) background metric. We now define
\begin{equation}
\psi_{ij} = \sqrt{f} \gamma_{ij}
\end{equation}
Since $f$ is ``slowly varying,'' to a good approximation, we have
\begin{equation}
\fl \int f t^{(0)}_{ab} t^a t^b = \frac{1}{32 \pi} \int d^4 x  \left[ \partial_i \psi_{jk} \partial^i \psi^{jk}
- 2 \partial_j \psi_{ik} \partial^i \psi^{jk} +2 \partial_j {\psi_i}^i \partial_k \psi^{jk}
-\partial_i {\psi_j}^j \partial^i {\psi_k}^k \right]
\label{t06}
\end{equation}
Even though our Riemannian normal coordinates are not globally well defined on the actual
spacetime, since $\psi_{ij}$ has support in $\mathcal D$, we can pretend that 
the coordinates $x^i$ range from $-\infty$ to $+\infty$. 
Let $\widehat{\psi}_{ij}$ denote the spatial Fourier transform of $\psi_{ij}$, i.e,
\begin{equation}
\widehat{\psi}_{ij} (t,k) = \frac{1}{(2 \pi)^{3/2}} \int d^3x \exp(-i k_l x^l) \psi_{ij} (t,x)
\end{equation}
We decompose $\widehat{\psi}_{ij}$ into its scalar, vector, and tensor parts via
\begin{equation}
\widehat{\psi}_{ij} =\widehat{\sigma} k_i k_j - 2 \widehat{\phi} q_{ij} + 
2 k_{(i} \widehat{z}_{j)} + \widehat{s}_{ij}
\end{equation}
where $q_{ij}$ is the projection orthogonal to $k^i$ of the Euclidean metric on Fourier transform
space and ${\widehat z}_i k^i = {\widehat s}_{ij} k^i = {\widehat s}_i{}^i = 0$.
With this substitution, our formula for the effective energy density of backreaction becomes
\begin{equation}
\int f t^{(0)}_{ab} t^a t^b = \frac{1}{32 \pi} \int dt d^3k \, k^ik_i  \left[ |{\widehat s}_{jk}|^2 - 8  
|\widehat{\phi}|^2 \right]
\label{t07}
\end{equation}

The term involving $|{\widehat s}_{jk}|^2$ arising from the ``tensor part'' of $\psi_{ij}$ is
positive definite and corresponds to the usual formula for the effective energy density
of short wavelength gravitational radiation~\cite{MTW}. This term can be ``large,''
corresponding to the well known fact that gravitational radiation can
produce large backreaction effects. In a cosmological context, this will contribute effects
equivalent to that of a $P = \rho/3$ fluid.
The term of potentially much greater interest for the
backreaction effects in cosmology associated with density inhomogeneities is the one involving
$|\widehat{\phi}|^2$, which arises from the ``scalar part'' of $\psi_{ij}$. This term is negative definite. The final---and
most difficult---step of the proof is to show that, in fact, this term is negligibly small.

In position space, the term of interest takes the form
\begin{equation}
E_\phi = - \frac{1}{4 \pi} \int d^4x \partial_i \phi \partial^i \phi
\label{Ep1}
\end{equation}
where $\phi$ is the inverse Fourier transform of $\widehat \phi$. This is of the form of Newtonian
potential energy\footnote{Note that we have {\em not} made any Newtonian approximations.
Note also that this formula is off by a factor of two from the standard 
formula for Newtonian gravitational energy.}. Furthermore, it follows from Einstein's equation
(\ref{ee1}) that $\phi$ satisfies a Poisson-like equation. To illustrate the basic idea of
our demonstration that $E_\phi$ is negligible, suppose that $\phi$ exactly satisfied the
Poisson equation
\begin{equation}
\partial^i \partial_i \phi = 4 \pi \sqrt{f} \rho
\label{poisson}
\end{equation}
[with $f$ as in eq.~(\ref{f})] and suppose it were known that $|\phi| \ll \sqrt{f}$ (as would be
expected since $|\psi_{ij}| = |\sqrt{f} \gamma_{ij}| \ll \sqrt{f}$). In that case, 
by integrating eq.~(\ref{Ep1}) by parts, we obtain
\begin{equation}
E_\phi =  \frac{1}{4 \pi} \int d^4x \phi  \partial_i  \partial^i \phi 
= \int d^4x \phi \sqrt{f}  \rho
\end{equation}
and hence
\begin{equation}
|E_\phi| \leq  \int d^4x |\phi| \sqrt{f}  \rho \ll \int d^4x f  \rho \sim \int d^4x f  \rho^{(0)} 
\sim \frac{1}{R^2} \int d^4x f
\end{equation}
which shows that $E_\phi$ indeed contributes
negligibly to the effective energy density. Here, as in eq.~(\ref{wec2}), the positivity of
$\rho$ was used to omit the absolute value sign on $\rho$ in the first inequality, and the
slowly varying character of $f$ was then used to replace $\rho$ by $\rho^{(0)}$. 

The actual proof that $E_\phi$ is negligible is much more difficult than as just sketched above
because (i) $\phi$ does not satisfy the simple Poisson equation (\ref{poisson}) but rather
an equation that contains many other terms that, a priori, are not negligibly small and (ii)
since $\phi$ is nonlocally related to $\psi_{ij}$, it is not obvious that $\phi$ is ``small'' in the required
sense, i.e., this must be shown. The reader wishing to see the details of how these
difficulties are overcome should read section II of our original paper~\cite{Green:2010qy}.
However, the key element of the proof is the argument sketched in the previous paragraph.

The above results show that, assuming only that (\ref{gam1}), (\ref{gam2}), (\ref{wec}), and (\ref{avg}) hold,
then in the absence of gravitational radiation, we must have $|t^{(0)}_{ab}| \ll 1/R^2$, i.e., the backreaction
effects effects of density inhomogeneities must be ``small.'' However, in the present era of precision cosmology,
it is of interest to know more precisely how ``small'' the backreaction effects are. In particular, what are the size
and nature of the various ``small corrections'' that we neglected in our analysis above
to the expansion rate and acceleration of the universe? Corrections as small as, say, $1\%$ would be of 
significant observational interest.

In order to analyze this, it is necessary to make further assumptions
about the nature of the stress-energy, $T_{ab}$, of matter and the
perturbed metric $\gamma_{ab}$. We assume that $T_{ab}$ takes the form
of a pressureless fluid, $T_{ab} = \rho u_a u_b$, and that appropriate
quasi-Newtonian behavior holds for both $T_{ab}$ and
$\gamma_{ab}$. With these assumptions, it is possible to solve
Einstein's equation to the accuracy required to compute the dominant
contributions to the terms that were neglected in the above
calculations. These calculations are quite involved, and we refer the
reader to~\cite{Green:2011wc} for all details (see, particularly,
Appendix B of that reference).  The upshot of these calculations is
that backreaction effectively modifies the matter stress-energy by
adding in the effects of kinetic motion of the matter as well as its
Newtonian potential energy and stresses\footnote{Note that for
  virialized systems, the kinetic motion and Newtonian potential
  contributions to stress cancel, so a universe filled with virialized
  systems behaves like a dust-filled universe~\cite{Baumann:2010tm}.}.
Consequently, for a quasi-Newtonian universe like ours appears to be,
the backreaction effects of small scale density inhomogeneities are
extremely small (far smaller than $1\%$), mainly involving only a
small ``renormalization'' of the mass density to take account of the
kinetic and Newtonian potential energy of matter.  This result is
complete agreement with the analysis of~\cite{Baumann:2010tm}, which
was done prior to our work~\cite{Green:2011wc}.

\section{Relationship to Our Mathematically Precise Formulation}

How can one make the arguments of the previous two sections more mathematically
precise and rigorous, so that the results can be stated as mathematical theorems
rather than heuristic estimates? Our approximations will become exact in the limit
that both $\gamma_{ab} \to 0$ and $L \to 0$, where $L$ denotes the homogeneity length
introduced in section I. Thus, if we wish to try to make these arguments mathematically precise,
we are led to consider a one-parameter family of metrics $g_{ab}(\lambda)$ and 
stress-energy tensors $T_{ab}(\lambda)$ such that, as $\lambda \to 0$, 
we have $g_{ab}(\lambda) \to g^{(0)}_{ab}$ (say, uniformly on
compact sets) and such that the homogeneity length $L \to 0$. Now, as $L \to 0$
there is no longer any need for the ``averaging field,'' $f^{ab}$, of eq.~(\ref{avg})
to be ``slowly varying,'' since everything is ``slowly varying'' as
compared with an arbitrarily small $L$. Thus, as $\lambda \to 0$, eq.~(\ref{avg}) 
becomes the statement that for any smooth tensor field $f^{ab}$ of compact support
we have 
\begin{equation}
\int f^{ab} \Delta T_{ab} \to 0 \, .
\label{avg2}
\end{equation}
Mathematically, eq.~(\ref{avg2}) is precisely the statement that the {\em weak limit}
as $\lambda \to 0$ of $\Delta T_{ab} (\lambda)$ vanishes.
However, note that we definitely do {\em not} want to require that 
$\Delta T_{ab} \to 0$ in a pointwise or uniform sense or we would be 
``throwing out the baby with the bathwater;'' we must allow the small scale inhomogeneities
to remain present as $\lambda \to 0$ so that we can see their possible backreaction effects. 

If we take the weak limit as $\lambda \to 0$ of Einstein's equation (\ref{ee2}) under the assumptions that
$\gamma_{ab} \to 0$ uniformly and $\Delta T_{ab} \to 0$ weakly, and 
if we also assume that $|\nabla_c \gamma_{ab}|$ remains 
bounded as $\lambda \to 0$, we find that
\begin{equation}
t^{(0)}_{ab} = -  \frac{1}{8 \pi} \wlim_{\lambda \to 0} G^{(2)}_{ab}
\label{wl1}
\end{equation}
which effectively replaces eq.~(\ref{t02}). Thus, for the one-parameter families that we wish to consider in order
to give a precise mathematical formulation of our results, the weak limit of the particular quadratic expression
in $\nabla_c \gamma_{ab}$ that appears on the right side of (\ref{wl1}) must exist. Note that it
is essential for our analysis that this quantity be allowed to be nonzero, since, otherwise,
we would preclude backreaction. It is mathematically convenient to slightly further restrict
the one-parameter
families we consider to require that
{\em all} quadratic expressions in $\nabla_c \gamma_{ab}$ have a well defined weak limit.

The above considerations lead us to the following
framework~\cite{Burnett:1989gp} for stating our results in a
mathematically precise form: We consider a one-parameter family of
metrics $g_{ab}(\lambda)$ and stress-energy tensors $T_{ab}(\lambda)$
such that the following conditions hold: (i) Einstein's equation
(\ref{ee1}) holds with $T_{ab}$ satisfying the weak energy
condition. (ii) $|\gamma_{ab}(\lambda)| \leq \lambda C_1(x)$ (where
$\gamma_{ab} (\lambda) \equiv g_{ab} (\lambda)- g^{(0)}_{ab}$) for
some positive function $C_1(x)$, so, in particular, $g_{ab}(\lambda)
\to g^{(0)}_{ab}$ uniformly on compact sets as $\lambda \to 0$.  (iii)
$|\nabla_c \gamma_{ab}(\lambda)| \leq C_2(x)$ for some positive
function $C_2(x)$, so derivatives of $\gamma_{ab}$ remain bounded as
$\lambda \to 0$. (iv) The weak limit of $\nabla_c \gamma_{ab}\nabla_d
\gamma_{ef}$ exists as $\lambda \to 0$.  This is precisely the
mathematical framework of our original
papers~\cite{Green:2010qy,Green:2011wc,Green:2013yua,Green:2014aga}. It
can be readily seen that condition (i) is precisely eqs.~(\ref{ee1})
and (\ref{wec}), condition (ii) is a precise version of (\ref{gam1}),
condition (iii) is a slightly strengthened version of (\ref{gam2}),
and condition (iv) corresponds (under Einstein's equation) to a
slightly strenthened version of (\ref{avg}). With the replacement of
(\ref{gam1}), (\ref{gam2}), (\ref{wec}), (\ref{avg}), and (\ref{ee1})
by conditions (i)--(iv), our heuristic arguments of the previous two
sections concerning the properties of $t^{(0)}_{ab}$ can be
transformed into mathematically precise theorems.

\section{Further Implications}

We have discussed above the application of our work to the analysis of backreaction effects
in cosmology. However, we believe that our work provides an indication of aspects
of Einstein's equation that may underlie fundamental stability properties of its solutions. 

We have derived, in a very general context, what may be viewed as the
``long wavelength effective equations of motion'' for the metric in
the presence of ``short wavelength disturbances.''  The key point is
that the long wavelength effective stress-energy tensor,
$t^{(0)}_{ab}$, associated with the short wavelength disturbances
always has positive energy properties\footnote{We proved that
  $t^{(0)}_{ab}$ satisfies the weak energy condition. We conjecture
  that $t^{(0)}_{ab}$ satisfies the dominant energy condition.},
provided only that the matter itself has positive energy. But
positivity of energy together with local conservation of total (i.e.,
real plus effective) stress-energy at long wavelengths suggests that
there cannot be a rapid, uncontrolled growth in solutions at long
wavelengths arising from the short wavelength behavior. In other
words, the nonlinear effects resulting from the insertion of a
perturbation at short wavelengths should not be able to
locally\footnote[1]{In asymptotically anti--de Sitter spacetimes,
  reflections off of $\mathscr{I}$ can lead to inverse
  cascades~\cite{Carrasco:2012nf}.} trigger a catastrophic ``inverse
cascade'' that has a large effect on the long wavelength behavior.
Although it is normally taken for granted that ``unphysical behavior''
of this sort does occur, it is a nontrivial feature for long
wavelength behavior to be ``protected'' in this manner from dynamical
effects occurring at short wavelengths.  Einstein's equation appears
to have this property. It is far from obvious that, e.g., various
modified theories of gravity will share this property.

\ack

We thank David Garfinkle and Luis Lehner for providing helpful
comments on our manuscript.  This research was supported in part by
NSF grants PHY-1202718 and PHY-1505124 to the University of
Chicago. This research was supported in part by Perimeter Institute
for Theoretical Physics. Research at Perimeter Institute is supported
by the Government of Canada through Industry Canada and by the
Province of Ontario through the Ministry of Research and Innovation.

\bibliography{mybib}

\end{document}